\documentclass[aps,pra,twocolumn,amsmath,amssymb,nofootinbib,superscriptaddress]{revtex4}
\usepackage[utf8]{inputenc}
\usepackage[T1]{fontenc}
\usepackage{microtype}
\usepackage[dvips]{graphicx}
\usepackage{xcolor}
\usepackage{times}
\usepackage{appendix}
\usepackage{chngcntr}
\usepackage{etoolbox}
\usepackage{lipsum}\usepackage{indentfirst} %wciecia po nazwach rozdzialow
\usepackage{amsmath}
\usepackage{amsfonts}
\usepackage{amssymb}
\usepackage{color}
\usepackage{graphicx}
\usepackage{caption}
\usepackage[version=3]{mhchem}
\usepackage{mwe}
\graphicspath{{figures/}}
\usepackage[subpreambles=true]{standalone}
\usepackage{import}

\usepackage[normalem]{ulem}
\usepackage{color}

\newcommand{\bra}[1]{\langle #1|}
\newcommand{\ket}[1]{|#1\rangle}

\newcommand{\be}{\begin{equation}}
\newcommand{\ee}{\end{equation}}
\newcommand{\bea}{\begin{eqnarray}}
\newcommand{\eea}{\end{eqnarray}}

\DeclareMathOperator{\dd}{dd}
\DeclareMathOperator{\sr}{sr}

\DeclareMathOperator{\osc}{osc}
\DeclareMathOperator{\erfc}{Erfc}
\DeclareMathOperator{\eff}{eff}
\DeclareMathOperator{\tf}{TF}
\DeclareMathOperator{\gpe}{GPE}

\usepackage[T1]{fontenc}

\begin{document}

\title{Breakdown of the mean field for dark solitons of dipolar bosons in a one-dimensional harmonic trap}
\author{Micha{\l}~Kowalski}
\affiliation{Center for Theoretical Physics, Polish Academy of Sciences, Al. Lotnik\'{o}w 32/46, 02-668 Warsaw, Poland}
\author{Rafa{\l}~O{\l}dziejewski}
\affiliation{Center for Theoretical Physics, Polish Academy of Sciences, Al. Lotnik\'{o}w 32/46, 02-668 Warsaw, Poland}
\author{Kazimierz~Rz\k{a}\.{z}ewski}
\affiliation{Center for Theoretical Physics, Polish Academy of Sciences, Al. Lotnik\'{o}w 32/46, 02-668 Warsaw, Poland}

%\date{\today}

\begin{abstract}
We directly compare the mean-field and the many-body approach in a one-dimensional Bose system in a harmonic trap. Both contact and dipolar interactions are considered. We propose a multi-atom version of the phase imprinting method to generate dark solitons in the system.  We begin with a general analysis of system dynamics and observe the emergence of a dark soliton and a shock wave. Center of mass and soliton motion become decoupled because the shock wave oscillates with the trap frequency and soliton does not. A detailed investigation of frequencies reveals significant differences between results obtained in the mean-field and the many-body pictures.

\end{abstract}

%\pacs{34.50.-s, 05.30.Jp, 67.85.-d}

\maketitle
\section{Introduction}
Solitons -- solutions of non-linear integrable differential equations which propagate without dispersion -- appear across many areas of science that range from physics to biology and medicine~\cite{dauxois2006}. Among a number of known equations supporting solitonic solutions an important example constitutes of the non-linear Schrodinger equation called also the Gross-Pitaevski equation (GPE). Although mainly used in the context of weakly interacting ultracold bosons~\cite{frantzeskakis2010dark}, it also describes the properties of the electric field of light in the non-linear media~\cite{kivshar1998dark}.\\
\indent If atoms in a Bose-Einstein condensate repel each other by point-like interactions, a resulting solitonic solution of the GPE takes on form of a dark soliton – a density dip with a phase jump across its density minimum. An analytical expression describing dark solitons in a uniform gas characterized by the GPE were already found in the 70s~\cite{zakharov71,zakharov73}. Shortly after achieving the first condensation in 1995, dark solitons were successfully generated by the phase imprinting method in the experimental setups with ultracold \ce{^{87}Rb}~\cite{burger1999dark} and \ce{^{23}Na}~\cite{denschlag2000generating} atoms trapped in a harmonic confinement.\\
\indent One of the most classic results about dark solitons in a BEC interacting only by short-range forces concerns its dynamics in a harmonic trap. In the so-called Thomas-Fermi regime for ultracold bosons, a characteristic frequency of soliton's oscillations is expressed by the trapping frequency $\omega$ and equals $\omega/\sqrt{2}$~\cite{busch2000motion} that was also observed in the experiment~\cite{weller2008}. However, this remarkably robust results for short-range interactions, does not hold for the dipole-dipole atomic interactions~\cite{bland2017}. In this case, the oscillation frequency of solitonic structures in a repulsive dipolar BEC -- predicted only in 2015~\cite{pawlowski2015dipolar,bland2015} -- highly depends on the strength of the atomic interactions and the interplay between local and non-local contributions to the total energy. With the recent progress on quantum gases consisting of atoms with considerable magnetic moments like \ce{^{164}Dy}~\cite{lu2011,maier2014} and Er~\cite{aikawa2012}, dipolar systems are now within experimental reach and call for deeper analysis.\\
\indent The non-linear mean-field (MF) theory of ultracold bosons provides an approximate description of interacting cold atoms. The underlying many-body (MB) model is linear and the state of the system is given by the many- body wave-function depending on positions of all particles. The discussion of correspondence between dark solitons present in the MF and many-body solutions of the full Hamiltonian has a long history. The best known example refers to the link between dark solitons moving on the circumference of a ring and type-II excitations from the seminal Lieb-Liniger model~\cite{LiebLiniger1963,Lieb1963} in the context of contact interacting particles, see~\cite{Kulish1976,ishikawa1980, KanamotoCarr2008,martin2010prl,Fialko2012,Sato2012,syrwid,Syrwid2016,Katsimiga2017bent,Brand2019,Kaminishi2019} and references therein. However, little is known about many-body states possesing features of dark solitons in a one-dimensional harmonic trap both for contact and purely dipolar interactions. Here, we aim to partially bridge this gap by studying many-body dark solitons in weakly interacting trapped systems of only few atoms far beyond the Thomas-Fermi regime. In particular, we investigate the oscillations of the many-body solitons and compare them to the dark solitons described by the GPE. Our results not only establish a link between the MF approach and the full many-body theory, but also can be verified in modern experiments with a precise control over only a few atoms in optical lattices or single traps (see for instance~\cite{greiner2002,serwane2011,meinert2015,baier2016,baier2018}).\\
\indent The work is organized as follows. In Sec.~\ref{model}, we introduce our many-body model for atoms interacting repulsively by contact or purely dipolar forces in a quasi-1D harmonic trap. We also remind the corresponding GPE. Then, in Sec.~\ref{solutions}, we analyse the many-body eigenstates of the system. Surprisingly, there is no good candidate for a many-body solitonic state among them, in a stark contrast to the ring geometry. Therefore, we introduce the many-body phase imprinting method of creating dark solitons in a harmonic trap. In Sec.~\ref{results}, we finally present our results. We investigate dynamics of the many-body solitons. We calculate the oscillations frequencies for different coupling strengths for both types of interactions and compare our findings with the GPE. The most important result is a significant disagreement between frequencies obtained in the MB and the MF approaches.
\section{The Model}\label{model}

We investigate a system of N repulsive bosons trapped in a harmonic potential
\begin{eqnarray}
U(x, y, z) = \frac{1}{2}m\omega^2 x^2 + \frac{1}{2}m\omega^2_\bot(y^2+z^2).
\end{eqnarray}
We assume that the transverse confinement is tight, so the wave function stays in the lowest energy level in the Y- and Z-directions meaning our system is quasi-1D. The aim of our study is to compare contact and dipolar solitons in many-body and mean-field approaches. 
The system of N repulsive bosons is often approximated by the using the quasi-1D GPE
\begin{subequations}
\begin{eqnarray}
\begin{split}
i\frac{\partial \psi_{\gpe}(\mathbf{r},t)}{\partial t} =  \psi_{\gpe}(\mathbf{r},t)\Big(-\frac{1}{2}\frac{\partial^2 }{\partial x^2}+U(\mathbf{r})+\\+\left(N-1\right)\int d\mathbf{r'} \lvert \psi_{\gpe}(\mathbf{r},t)\rvert^2 V(\mathbf{r-r'})\Big),
\end{split}\\
\int d\mathbf{r}\lvert \psi_{\gpe}(\mathbf{r},t)\rvert^2 = N
\end{eqnarray}\label{eq:gpe}
\end{subequations}
where $U(\mathbf{r})$ is the trapping potential and $V(\mathbf{r})$ is an interaction potential. Throughout the paper we use oscillatory units with $\hbar \omega$, $\sqrt{\frac{\hbar}{m\omega}}$, $\sqrt{\hbar m \omega}$, $\frac{1}{\omega}$ as units of energy, length, momentum and time respectively.  In the context of ultracold atoms, this equation is also called the MF description of weakly interacting bosons. This approach provided correct predictions of many properties of the Bose-Einstein condensate, including its shape, energy, normal modes of excitations and many other nonlinear phenomena. Moreover, the GPE supports dark solitons in a form of solutions with a density notch and a quickly changing phase, which have also been experimentally produced with the phase imprinting method. However, the MF model assures a simplified description of system of N repulsive bosons based on a naive assumption that every atom is in the same state. It is only an approximation of the more fundamental many-body approach. As the nonlinear GPE supports solitons, it is important to look for solitons in the linear many-body approach and compare them. It has been done in the Lieb-Liniger model~\cite{LiebLiniger1963,Lieb1963} but, to the best of our knowledge, this is the first paper where multi-atomic solitons are considered for the harmonically trapped system.

In order to describe a system of N bosons following the many-body approach, one needs to derive the wave function depending on positions of all particles. One of possible ways of deriving the many-body wave function is to diagonalize a Hamiltonian matrix. The Hamiltonian of the system under investigation can be written in a form:
\begin{eqnarray}\label{eq:ham0}
H=\sum_i^N T_i + \sum_i^N \frac{1}{2} x_i^2 + \sum_{i<j}^NV_{ij},
\end{eqnarray}
with $T_i$ being the single-particle kinetic energy operator and $V_{ij}$ the two-body interaction operator. The Hamiltonian \eqref{eq:ham0} can be rewritten as a sum of the Hamiltonian of the noninteracting quantum harmonic oscillator and the interaction term. Therefore, in the second-quantization the Hamiltonian \eqref{eq:ham0} reads:
%where $a_j and a_j^\dagger$ are bosonic annihilation and construction operators
\begin{eqnarray}\label{eq:hamiltonian}
H= H_{\osc} + \frac{1}{2}\int\hat{\Psi}^\dagger(x)\hat{\Psi}^\dagger(y)V(x-y)\hat{\Psi}(x)\hat{\Psi}(y)dxdy,
\end{eqnarray}
with $\hat{\Psi(x)}$ being a bosonic field operator and $H_{\osc} =\frac{1}{2} \int\hat{\Psi}^\dagger(x)(-\frac{\partial^2}{\partial x^2} +x^2)\hat{\Psi}(x)dx$. We study systems where V(r) is either a short range $V_{\sr}(r)$ or a long range dipolar $V_{\dd}(r)$ potential. The short range potential is $V_{\sr}(r)=g_{\sr}\delta(r)$ with parameter $g_{\sr} = \int V_{\sr}(r) dr$ defining strength of the interaction. We study repulsive systems with $g>0$. In order to obtain the explicit formula describing $V_{\dd}$, we follow the procedure described in paper \cite{pawlowski2015dipolar}. \\
We introduce an aspect ratio of the trap $\sigma = \frac{\omega_{\perp}}{\omega}$ and a dipolar coupling strength $A_{\dd}$ yielding 
\begin{eqnarray}\label{v_int}
V_{\dd}(r) = \frac{A_{\dd}}{\sigma^2} \frac{1}{\sigma} V_{\eff}(\frac{r}{\sigma})
\end{eqnarray}
with a term 
\begin{eqnarray}
V_{\eff}(u)=\frac{3}{4}[-2|u| + \sqrt{2\pi}(1+u^2)e^{u^2/2} \erfc (\frac{|u|}{\sqrt{2}}).
\end{eqnarray}
This effective quasi-1D potential comes from the integration of the full 3D dipolar interaction over both transverse variables. The area of our interest is a long range part of interaction, so we assume that the contact term of the effective dipolar potential is exactly cancelled possibly with the help of Feshbach resonances.
As we want to investigate similarities and differences between systems interacting via contact and dipolar forces, we define dipolar strength parameter $g_{\dd}$
\begin{eqnarray}
g_{\dd} = \int V_{\dd}(r) dr = \frac{3 A_{\dd}}{\sigma^2}.
\end{eqnarray}
From now on, we will keep $\sigma = 0.1$ and compare systems described by the same strength parameters $g_{\dd} = g_{\sr} = g$. It is important to mention that in the case of the MF approach only a gas parameter $(N-1)g$ defines the system. However, in the multi-atom approach both $N$ and $g$ are separately relevant.

From the numerical point of view, the most optimal basis to describe the system are quantum harmonic oscillator eigenstates. We use the second quantization formalism and define a basis of Fock states. We take into account all states with a given number of particles, which energies are smaller than a cut-off energy. Defining the cut-off in the energy space, rather than in the momentum space is more efficient method of obtaining numerical convergence \cite{sowinski}.

In order to obtain eigenstates and energies of the system, we diagonalize the Hamiltonian matrix
\begin{eqnarray}
H_{ij}=\bra{i}\hat{H}\ket{j},
\end{eqnarray}
where $\ket{i}, \ket{j}$ are states belonging to the Fock space.
%It is worth noting that Fock space can be split into two subspaces depending on symmetry of wave functions (odd or even) described by given state. (O tej symetrii chcę wspomnieć przy nadruku, dlatego wspominam)

%UWAGA! może ta sekcja solutions wejdzie w sekcje model, jeszcze nie wiem jak rozpiszesz model, zobaczymy
\section{Solutions}\label{solutions}
Having access to both eigenenergies and eigenstates we are ready to look for solitons in the system. Following recent papers investigating many-body solitons in the Lieb-Liniger model \cite{syrwid,oldziejewski2018a}, one could think that also in this case dark solitons could be identified among eigenstates of the Hamiltonian \eqref{eq:hamiltonian}. \\Studying the repulsive case, we are interested in properties of dark solitons. In this situation density forms a single notch and in the area of the notch phase exhibits a jump of $\pi$. Keeping this in mind, one can ask a question if any single particle state fulfil these conditions.

We start our analysis with the ideal gas. In this case, the excited state of quantum harmonic oscillator (QHO),$\bigotimes^N \ket{0,1,0,...} = \ket{0,N,0,...}$, seems like a reasonable candidate because it has both the density dip and the phase jump, so one can try to find the eigenstate of the interacting system with the highest contribution of the aforementioned state among all the Fock states.

This turnes out to be non-trivial even for weak interactions. We expected the maximum $\ket{0,N,0,...}$ occupation tend to one as the interactions become weaker. Instead, it was approaching different values depending on a number of particles considered. This is a peculiar property of the harmonic trap caused by evenly spaced energies of $H_{\osc}$. Once the interactions become weaker, the energy of the eigenstate with the highest $\ket{0,N,0,...}$ contribution approaches $\frac{3N}{2}$ but there are multiple other states with the same energy leading to degeneracy. For example in the case of N=2, $\ket{0,2,0}$ has the same energy as $\ket{1,0,1}$ namely 3. Hence in contrast with the Lieb-Liniger model, these eigenstates remain the combination of several other states with the energy equal to $\frac{3N}{2}$ even for vanishing interactions.

Even if the excited eigenstate would form the dark soliton, still it would be hard to realize this state in the experiment. Therefore we decided to follow a different approach and try to replicate the experimental procedure of phase imprinting. This method creates a dark soliton in a BEC via a pulse of a far detuned laser applied on one half of the condensate and so creates a phase difference between the left and the right side. The length of this pulse is tuned to create a phase difference of $\pi$ and hence causes emergence of the dark soliton. There are not many papers discussing the phase-imprinting method in the many-body approach \cite{schmelcher2015many}. To the best of our knowledge it was only applied together with density engineering, which is not the case in the experimental realisation. As in real life, our implementation of phase imprinting modifies only the phase of the wave function and is equivalent to multiplying the ground state wave function by an arbitrary phase factor
\begin{eqnarray}
\Psi(x_1,x_2,...,x_N)=\Phi(x_1,x_2,...,x_N)e^{i\phi(x_1,x_2,...x_N)},
\end{eqnarray}
where $\Psi$ is the many-body wave function of a solitonic state, $\Phi$ is the ground state wave function and $\phi$ is an arbitrary phase factor. For the numerical convenience we choose $\phi(x_1,x_2,...x_N) = \sum_j^N \tan^{-1}(\alpha x_j)$, where $\alpha$ is the parameter changing sharpness of phase jump which can also be controlled in experiments. The optimal sharpness of the phase jump has to be tuned to fit healing length of the soliton. It means that the stronger the interaction the narrower the phase jump has to be.
%\begin{eqnarray}
%\Psi(x_1,x_2,...,x_N)=\Phi(x_1,x_2,...,x_N)e^{i*\sum_j^N tan^{-1}(\alpha x_j)},
%\end{eqnarray}
As for now, the solitonic wave function $\Psi$ is merely an initial condition. We derive the time evolution of the system by expressing $\Psi$ in the basis of eigenstates of the system as follows:
\begin{eqnarray}
\Psi(x_1,x_2,...,x_N, t) = \sum _{i} \beta_{i} \psi_{i}( x_{1} ,...,x_{N}) \exp\left( -iE_{i} t\right)\\
\beta _{i} =\int dx_{1} ...dx_{N} \psi ^{*}_{i}( x_{1} ,...,x_{N}) \Psi ( x_{1} ,...,x_{N}),
\end{eqnarray}
where $\psi_{i}( x_{1} ,...,x_{N})$ and $E_i$ are the eigenstates and the eigenvalues of the Hamiltonian \eqref{eq:hamiltonian}.
In order to visualise a soliton, we derive a one-particle density
\begin{eqnarray}
\rho (x_1,t) =\int dx_{2} ...dx_{N} \Psi^* ( x_{1} ,...,x_{N} ,t)\Psi(x_1,x_2,...,x_N, t).
\end{eqnarray}
%The fact that this approach works is another difference between solitons in a harmonic trap and in Lieb-Liniger model. In the case of atoms on the ring the system is translationally invariant, hence the one-particle density would be uniform. It is not the case in the harmonic trap or in the box with hard walls, meaning that one-particle density reveals solitons in those systems.

As we aim to obtain the MF dark solitons from Eq. \eqref{eq:gpe} and compare them with MB solutions, we employ an analogous scheme. Firstly, we find a ground state $\psi_{\gpe}(x)$  for given parameters by using the well-known imaginary time evolution (ITE) technique. At this point, we can compare ground states obtained in MF and MB approaches. Both density profiles and ground-state energies (up to 2 \% difference for the highest $g$) are in a very good agreement for both dipolar and contact interactions and for all coupling strenghts considered in this work. Then, we imprint the same phase as in the MB calculations, namely $\Psi_{\gpe}(x)=\psi_{\gpe}(x)e^{i\phi_{MF}(x)}$, with $\phi_{MF}=\tan^{-1}(\alpha x)$. Finally, we evolve Eq. \eqref{eq:gpe} in a standard real-time evolution with $\Psi_{\gpe}(x)$ as an initial condition.

Note that we can calculate the quantum depletion for any many-body state, in particular for a many-body ground state before and after phase imprinting, by diagonalazing a single particle density matrix constructed from the many-body wave function. It provides a tool for comparing mean-field and many-body results.
\section{Results}\label{results}
\begin{center}
\begin{figure}[t]
\includegraphics[width=0.455\textwidth]{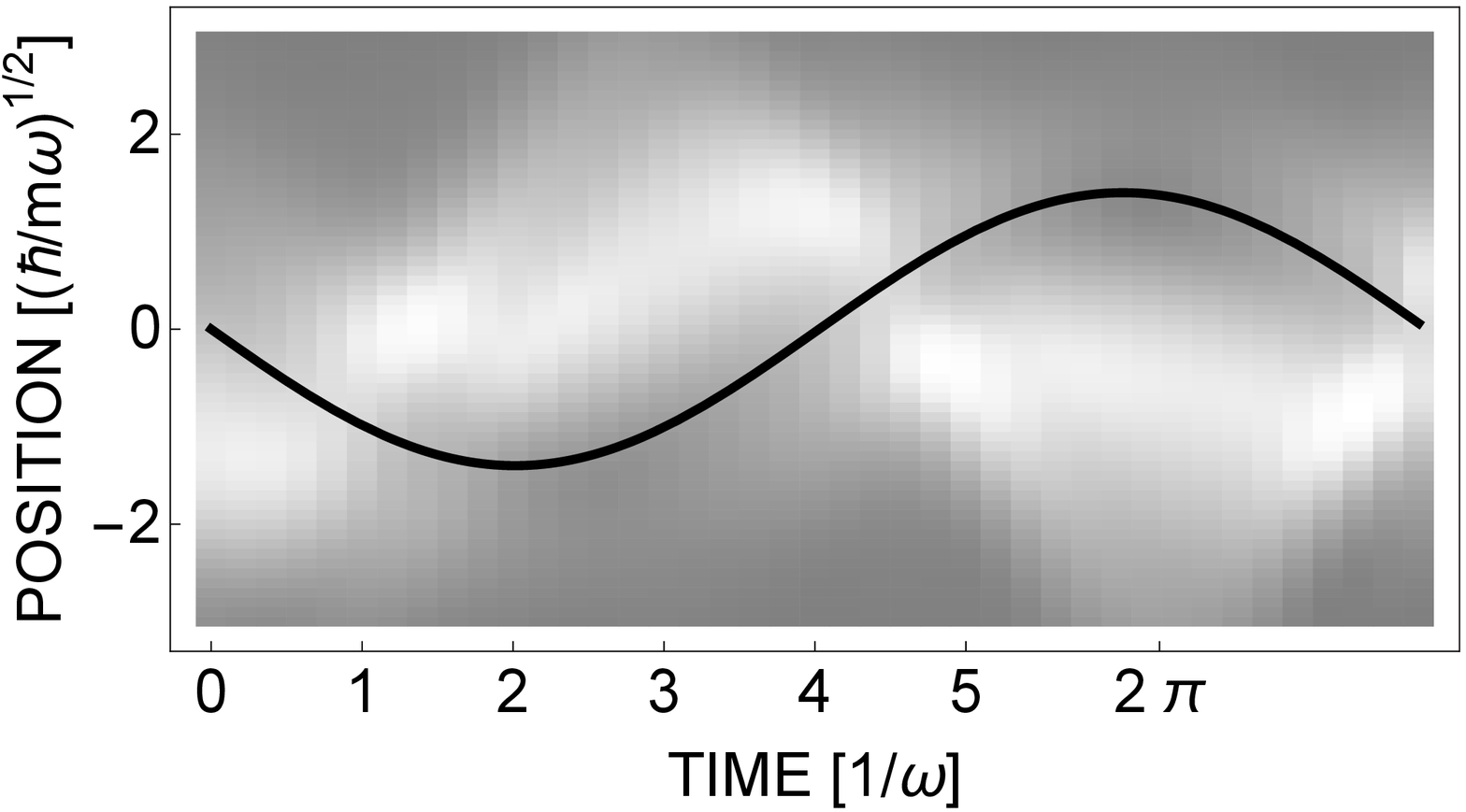}
    \caption{(color online) One-particle density evolution of a dipolar dark soliton (black solid line) and a shock wave (white trace) for $N=6$ particles and $g=0.3$. The dark soliton is oscillating with a frequency of $0.94\,\omega$ and the shock wave is oscillating with the frequency of the trap $\omega$.}\label{fig:evo}

\includegraphics[width=0.455\textwidth]{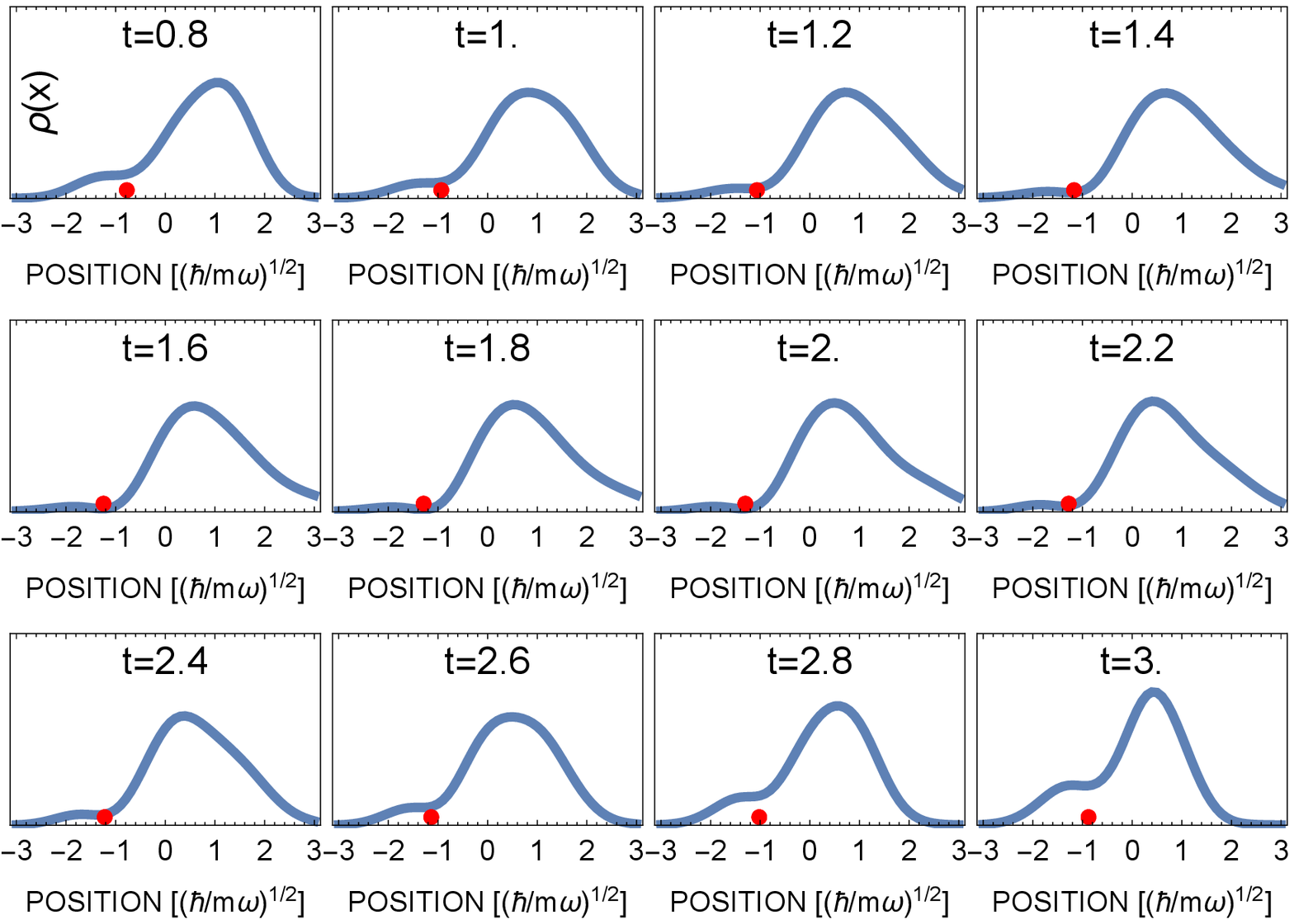}
    \caption{(color online) Sequence of images showing a spatial density profile in consecutive time-steps for the situation from Fig. \ref{fig:evo} ($N=6$ dipolar particles and $g=0.3$). Time $t$ is given in units of $\frac{1}{\omega}$. Red dot indicates the position of the soliton. The soliton becomes deeper until density reaches zero -- the soliton bounces from the trap and begins moving to the other side of the trap. Once moving towards the center of the trap it becomes shallower.}\label{fig:ani}
 \end{figure}
\end{center}

Having a model of the experimental method of phase imprinting and being able to calculate the evolution of dark solitons, we can focus on properties of contact and dipolar many-body solitons and compare them with the MF results. Firstly, we would like to focus our attention on general aspects of the evolution of many-body solitons in the harmonic trap. In order to study the evolution of the system  we plot the one-particle density as a function of time and space in Fig. \ref{fig:evo} for $N=6$ dipolar bosons and $g=0.3$. It reveals that the phase imprinting method causes not only the dark soliton to emerge but also a shock wave in the form of a density peak initially moving in the opposite direction to the soliton. Plots of density profiles in consecutive time steps shown in Fig. \ref{fig:ani} reveal more details of soliton evolution. The local density minimum moves from the center of the trap towards the left side as long as the density of the notch is greater than zero. When the soliton becomes black, namely when the one-particle density in the dip reaches zero, its velocity also equals zero, both indicating a turning point. The soliton begins to move right and becomes shallow in the center of the trap. The relation between the depth of the dark soliton and its velocity is one of the fundamental properties of solitons and have been studied in a number of papers \cite{zakharov71,zakharov73,parker2010dark}.

%\begin{center}
%\begin{figure}[h!]
%\includegraphics[width=0.455\textwidth]{s1.eps}
%    \caption{(color online) One-particle density showing evolution in time of a dipolar dark soliton and a shock wave. The dark soliton is moving with a frequency of $0.86 \omega_0$ and the shock wave is oscillating with the frequency of the trap $\omega_0$.}\label{fig:evo}
% \end{figure}
%\end{center}

As we pointed before the phase imprinting method creates a soliton but also a shock wave. This effect has been already observed in experiments implementing phase imprinting \cite{burger1999dark, denschlag2000generating}. Both the shock wave and the soliton oscillate in the trap harmonically %with different frequencies. This leads to decoupling of center of mass and soliton movement what has been previously revealed by mean-field solitons \cite{parker2010dark}
but the shock wave oscillates with the trap frequency. It is then worth to study the frequency of the soliton movement as it was one of the factors differentiating contact and dipolar solitons in the mean-field approach \cite{bland2017}.

One of properties of contact solitons revealed by number of studies \cite{fedichev1999, busch2000motion, parker2010dark} is that the frequency of oscillation does not depend on the strength of interactions and equals $\omega_{\tf} = \frac{1}{\sqrt{2}}\omega$. However, this result is obtained in the Thomas-Fermi (TF) limit assuming that the background density varies slowly on the scale of the soliton. The kinetic energy of particles in the TF limit can be neglected compared to the potential energy. However, satisfying this condition in the case of small systems would demand very strong interactions causing atoms to deplete the ground state and thus making our and the mean-field result incomparable. Our studies focused on the case of small systems in the many-body approach far from the TF limit and with the quantum depletion of the ground state before phase imprinting not exceeding 5\%. 

In order to analyse the frequency of oscillation, we trace the position of a local minimum of the one-particle density (in the many-body approach) or condensate wave function (in the mean-field picture) in consecutive time steps. We trace the minimum until it crosses the center of the trap which gives us half of the period. 
%We calculate the frequency of oscillations of contact dark solitons for systems of N=4, N=6 and N=8 particles in the range of gas parameters $Ng$ as shown in Fig. \ref{fig:cont}. It is important to notice that the frequency changes continuously and decreases monotonically with the strength of interaction still far from the TF limit.
%    \begin{figure}
 %       \includegraphics[width=0.455\textwidth]{c1.eps}
 %       \caption{(color online) Frequency of dark soliton oscillation as a function of gas parameter for different number of bosons. Mean-field result obtained in Thomas-Fermi limit is plotted with red line. As our studies are conducted far from T-F regime, the frequency for N=4, N=6 and N=8 decreases with gas parameter. Lines are fitted to points}
%        \label{fig:cont}
%    \end{figure}

Recent paper investigating dipolar solitons in the mean-field approach revealed significant differences between contact and dipolar solitons in the TF regime \cite{bland2017}, one of them exhibited by the frequency of oscillations. In contrast to previously described mean-field contact solitons, the frequency of dipolar solitons depends on the interaction strength and, in general, the dipolar soliton frequency is smaller than the one obtained for contact solitons. It is then worth asking if many-body solitons exhibit similar behaviour already for weak interactions. %In order to answer this question we plot frequency as the function of gas parameter, what can be seen in Fig. \ref{fig:dip}. Once again, the frequency decreases monotonically with the strength of interaction and for the same gas parameter is always smaller with the growing number of particles.
%\begin{center}
        %\begin{figure}
        %\includegraphics[width=0.455\textwidth]{d1.eps}
        %\caption{(color online) Frequency of dark soliton oscillation as a function of gas parameter %$g_{\dd}$ for dipolar interaction. With increasing number of particles frequency decreases.}
%        \label{fig:dip}
%    \end{figure}
%\end{center}

\begin{center}
    \begin{figure}
        \includegraphics[width=0.455\textwidth]{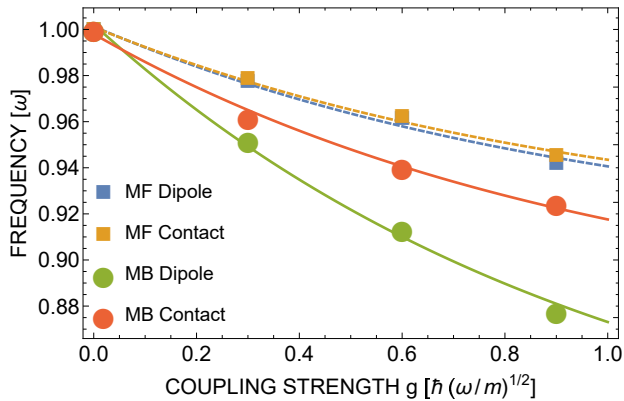}
        \caption{(color online) Comparison of frequency as a function of a coupling strength $g$ for contact and dipolar interactions in the many-body (circles) and mean-field (squares) approaches for $N=6$ particles. Mean-field solitons are almost indistinguishable with dipolar ones being only slightly slower. On the other hand, many-body solitons differ significantly. Dipolar solitons always oscillate slower as for the MF dark solitons in the TF regime studied in~\cite{bland2017}.}
        \label{fig:comp}
    \end{figure}
\end{center}

To answer this question we directly compare many-body and mean-field solitons for contact and dipolar interactions. We plot frequencies obtained for system of N=6 particles for the increasing coupling strength $g$ in Fig. \ref{fig:comp}. Firstly, we note that the frequencies of MF and MB solitons differ significantly. In the MF picture, contact and dipolar solitons are almost indistinguishable. In opposition, MB dipolar solitons oscillate much slower than contact ones which agrees with the recent MF analysis within the TF approximation in~\cite{bland2017}. 

It is then important to ask why the MF model far away from the TF regime is inconsistent when applied to the studied system. We can indicate two factors that may play a significant part. On the technical level, it seems like rapidly changing excited QHO states contribute to the difference in the energy between short-range and dipolar interactions in the many-body picture. On the other hand, the MF wave-function does not vary at the scale given by the range of dipolar interaction. Hence, for systems far from the TF regime, the dipolar interacting scenario almost does not differ from the short-range interaction case. The other factor contributing to the difference between MB and MF solutions is the phase imprinting method. Before imprinting, the systems are comparable as the depletion of the MB ground state does not exceed 5\%.  After the procedure, it rises by 10-20\% depending on the sharpness of the phase jump.  It means that the excited fraction cannot be neglected anymore in the MB calculations while it is not present at all in the MF case. This is a very important observation as the depletion of the ground state is fundamentally bound with phase-imprinting method and need to be taken into consideration when studying small systems both theoretically and experimentally.

Having discussed the differences between MF and MB results, we focus on the properties of MB contact and dipolar solitons. The system proves to be interaction-sensitive as the frequency varies significantly with both the strength and the range of the interaction. In both cases, the frequency decreases with the increasing coupling strength $g$. The dipolar solitons always oscillate slower than their short-range counterparts. As we are far from the TF limit, the frequency of contact solitons does not converge to $\frac{1}{\sqrt{2}}\omega$.

We have investigated not only the frequency of solitons oscillation but also their lifetimes. While it is hard to define a sharp condition for the soliton to be indistinguishable from the background, we noted that contact solitons live significantly longer than dipolar counterparts, with the lifetime strongly dependent on the coupling strength $g$.
%In the limit of weak interactions there is no correlation between width of the imprint namely parameter $\alpha$ and evolution of soliton because its healing length is larger than size of the trap. When the interaction becomes stronger transitional region of phase jump must be narrower. 

\section{Conclusions}\label{conclusions}
The goal of this paper is to compare dark solitons in the mean-field and many-body approaches for contact and dipolar interactions. We have began our many-body analysis with calculating eigenstates and energies of the system via the numerical diagonalization of the Hamiltonian matrix. We have found that one cannot identify the dark solitons with a specific eigenstate of the system, in stark contrast to the well-known situation of atoms in a ring trap. This follows directly from quantum degeneracy of multi-particle eigenstates of the non-interacting gas in the harmonic trap because the single particle eigenenergies are spaced evenly.

In order to study many-body solitons in the harmonic trap we have introduced the multi-atom version of the phase imprinting method. Just as in the classic experiment it causes not only the soliton but also the shock wave to appear. Those waves oscillate with different frequencies and thus movements of the soliton and the center of mass are decoupled. We investigate the frequency of oscillation of dark solitons to reveal similarities and differences between contact and dipolar solitons and compare our many-body analyses with mean-field results.

Although calculated for small and weakly interacting systems, our studies comparing many-body contact and dipolar solitons uncover similar features as previously discussed mean-field results within the Thomas-Fermi regime~\cite{bland2017}. The frequency of oscillations for dipolar solitons strongly depends on the coupling strength and is lower compared to contact solitons for the corresponding interaction strength. For comparison, we also analyzed contact and dipolar solitons in our small system induced by the phase imprinting method at the MF level. The MF approach fails in the case of our system as the dipolar and contact solitons are almost identical and their properties differ significantly from the MB solitons. We can define two factors that cause significant differences between our MB and MF solitons. Firstly, the quickly oscillating excited QHO states play an important role in the case of dipolar interaction. Secondly, the phase imprinting method enlarges a depletion of the ground state and the excited fraction is no longer negligible.
\begin{acknowledgments}
We thank K. Pawłowski for his careful and critical reading of the manuscript and fruitful discussions. This work was supported by the (Polish) National Science Center Grants 2016/21/N/ST2/03432 (RO and MK) and 2015/19/B/ST2/02820 (KR). Center for Theoretical Physics is a member of KL FAMO.
\end{acknowledgments}
\bibliographystyle{apsrev4-1}
\bibliography{artbib}
\end{document}